\documentclass[aps,apl,twocolumn,groupedaddress,amsmath]{revtex4}
\usepackage[latin1]{inputenc}    
\usepackage{graphicx}   
\usepackage{xspace}

\begin{document}
\def\imat  {i}
\def\be{\begin{equation}}
\def\ee{\end{equation}}
\def\bea{\begin{eqnarray}}
\def\eea{\end{eqnarray}}
\def\la{\langle}
\def\ra{\rangle}
\def\l{\langle\!\langle}
\def\r{\rangle\!\rangle}

\def\Gin{\Gamma_\mathrm{in}}
\def\Gout{\Gamma_\mathrm{out}}
\def\Gs{\Gamma_\mathrm{S}}
\def\Gd{\Gamma_\mathrm{D}}
\def\Gc{\Gamma_\mathrm{C}}
\def\Grel{\Gamma_\mathrm{rel}}
\def\Gabs{\Gamma_\mathrm{abs}}
\def\Gem{\Gamma_\mathrm{em}}
\def\Gdet{\Gamma_\mathrm{det}}
\def\tdet{\tau_\mathrm{det}}
\def\Vtdeg{V_\mathrm{2DEG}}
\def\Iqpc{I_\mathrm{QPC}}
\def\Vqpc{V_\mathrm{QPC}}
\def\Gqpc{G_\mathrm{QPC}}
\def\Vsd{V_\mathrm{SD}}
\def\Vgl{V_\mathrm{G1}}
\def\Vgr{V_\mathrm{G2}}
\def\aGl{\alpha_\mathrm{G1}}
\def\aGr{\alpha_\mathrm{G2}}
\def\Idqd{I_\mathrm{DQD}}
\def\tc{t_\mathrm{C}}
\def\mul{\mu_\mathrm{1}}
\def\mur{\mu_\mathrm{2}}
\def\mus{\mu_\mathrm{S}}
\def\mud{\mu_\mathrm{D}}
\def\Ec{E_\mathrm{C}}
\def\Ecm{E_\mathrm{Cm}}
\def\Ecl{E_\mathrm{C1}}
\def\Ecr{E_\mathrm{C2}}
\def\Vqsd{V_\mathrm{QPC-SD}}
\def\Vdsd{V_\mathrm{DQD-SD}}
\def\Rs{R_\mathrm{S}}
\def\Rf{R_\mathrm{F}}
\def\Cc{C_\mathrm{C}}
\def\Cf{C_\mathrm{F}}
\def\Vout{V_\mathrm{out}}
\def\Vn{V_\mathrm{n}}
\def\Vn{I_\mathrm{n}}
\def\fBW{f_\mathrm{BW}}


\newcommand*{\eg}{e.\,g.\xspace}
\newcommand*{\ie}{i.\,e.\xspace}
\newcommand*{\cf}{c.\,f.\,,\xspace}
\newcommand*{\etc}{etc.\@\xspace}
\newcommand*{\etal}{\textsl{et al.\@\xspace}}

\newcommand*{\e}{\mathrm{e}}       
\newcommand*{\ci}{\mathrm{i}}
\newcommand*{\kb}{\ensuremath{\,k_\mathrm{B}}\xspace}
\newcommand*{\abs}[1]{\mid#1\mid}


\newcommand*{\um}{\ensuremath{\,\mu\mathrm{m}}\xspace}
\newcommand*{\nm}{\ensuremath{\,\mathrm{nm}}\xspace}
\newcommand*{\mm}{\ensuremath{\,\mathrm{mm}}\xspace}
\newcommand*{\m}{\ensuremath{\,\mathrm{m}}\xspace}
\newcommand*{\sqm}{\ensuremath{\,\mathrm{m}^2}\xspace}
\newcommand*{\sqmm}{\ensuremath{\,\mathrm{mm}^2}\xspace}
\newcommand*{\squm}{\ensuremath{\,\mu\mathrm{m}^2}\xspace}
\newcommand*{\psqm}{\ensuremath{\,\mathrm{m}^{-2}}\xspace}
\newcommand*{\psqmV}{\ensuremath{\,\mathrm{m}^{-2}\mathrm{V}^{-1}}\xspace}
\newcommand*{\cm}{\ensuremath{\,\mathrm{cm}}\xspace}

\newcommand*{\nF}{\ensuremath{\,\mathrm{nF}}\xspace}
\newcommand*{\pF}{\ensuremath{\,\mathrm{pF}}\xspace}

\newcommand*{\emob}{\ensuremath{\,\mathrm{m}^2/\mathrm{V}\mathrm{s}}\xspace}
\newcommand*{\edos}{\ensuremath{\,\mu\mathrm{C}/\mathrm{cm}^2}\xspace}
\newcommand*{\mbar}{\ensuremath{\,\mathrm{mbar}}\xspace}

\newcommand*{\A}{\ensuremath{\,\mathrm{A}}\xspace}
\newcommand*{\nA}{\ensuremath{\,\mathrm{nA}}\xspace}
\newcommand*{\pA}{\ensuremath{\,\mathrm{pA}}\xspace}
\newcommand*{\fA}{\ensuremath{\,\mathrm{fA}}\xspace}
\newcommand*{\uA}{\ensuremath{\,\mu\mathrm{A}}\xspace}

\newcommand*{\Ohm}{\ensuremath{\,\Omega}\xspace}
\newcommand*{\kOhm}{\ensuremath{\,\mathrm{k}\Omega}\xspace}
\newcommand*{\MOhm}{\ensuremath{\,\mathrm{M}\Omega}\xspace}
\newcommand*{\GOhm}{\ensuremath{\,\mathrm{G}\Omega}\xspace}

\newcommand*{\Hz}{\ensuremath{\,\mathrm{Hz}}\xspace}
\newcommand*{\kHz}{\ensuremath{\,\mathrm{kHz}}\xspace}
\newcommand*{\MHz}{\ensuremath{\,\mathrm{MHz}}\xspace}
\newcommand*{\GHz}{\ensuremath{\,\mathrm{GHz}}\xspace}
\newcommand*{\THz}{\ensuremath{\,\mathrm{THz}}\xspace}

\newcommand*{\K}{\ensuremath{\,\mathrm{K}}\xspace}
\newcommand*{\mK}{\ensuremath{\,\mathrm{mK}}\xspace}

\newcommand*{\kV}{\ensuremath{\,\mathrm{kV}}\xspace}
\newcommand*{\V}{\ensuremath{\,\mathrm{V}}\xspace}
\newcommand*{\mV}{\ensuremath{\,\mathrm{mV}}\xspace}
\newcommand*{\uV}{\ensuremath{\,\mu\mathrm{V}}\xspace}
\newcommand*{\nV}{\ensuremath{\,\mathrm{nV}}\xspace}

\newcommand*{\eV}{\ensuremath{\,\mathrm{eV}}\xspace}
\newcommand*{\meV}{\ensuremath{\,\mathrm{meV}}\xspace}
\newcommand*{\ueV}{\ensuremath{\,\mu\mathrm{eV}}\xspace}

\newcommand*{\T}{\ensuremath{\,\mathrm{T}}\xspace}
\newcommand*{\mT}{\ensuremath{\,\mathrm{mT}}\xspace}
\newcommand*{\uT}{\ensuremath{\,\mu\mathrm{T}}\xspace}

\newcommand*{\ms}{\ensuremath{\,\mathrm{ms}}\xspace}
\newcommand*{\s}{\ensuremath{\,\mathrm{s}}\xspace}
\newcommand*{\us}{\ensuremath{\,\mathrm{\mu s}}\xspace}
\newcommand*{\rpm}{\ensuremath{\,\mathrm{rpm}}\xspace}
\newcommand*{\minute}{\ensuremath{\,\mathrm{min}}\xspace}
\newcommand*{\degree}{\ensuremath{\,^\circ\mathrm{C}}\xspace}

\newcommand*{\EqRef}[1]{Eq.~(\ref{#1})}
\newcommand*{\FigRef}[1]{Fig.~\ref{#1}}

 \title{Detecting single-electron tunneling involving virtual processes in real time}
 \author{S.~Gustavsson}
 \email{simongus@phys.ethz.ch}
 \author{M.~Studer}
 \author{R.~Leturcq}
 \author{T.~Ihn}
 \author{K.~Ensslin}
 \affiliation {Solid State Physics Laboratory, ETH Z\"urich, CH-8093 Z\"urich,
 Switzerland}
\author{D.~C. Driscoll}
\author{A.~C. Gossard}
\affiliation{Materials Departement, University of California, Santa
Barbara, CA-93106, USA}

\date{\today}

\begin{abstract}
We use time-resolved charge detection techniques to probe virtual
tunneling processes in a double quantum dot. The process involves an
energetically forbidden state separated by an energy $\delta$ from
the Fermi energy in the leads. The non-zero tunneling probability
can be interpreted as cotunneling, which occurs as a direct
consequence of time-energy uncertainty.
For small energy separation the electrons in the quantum dots
delocalize and form molecular states. In this regime we establish
the experimental equivalence between cotunneling and sequential
tunneling into molecular states for electron transport in a double
quantum dot.
Finally, we investigate inelastic cotunneling processes involving
excited states of the quantum dots. Using the time-resolved charge
detection techniques, we are able to extract the shot noise of the
current in the cotunneling regime.
\end{abstract}

\maketitle

\section{Introduction}
A semiconductor double quantum dot (DQD) is the mesoscopic analogue
of a diatomic molecule. The energy levels and the interdot coupling
energy can be precisely controlled with gate voltages
\cite{vanderwiel:2002}, which allows the DQD to be tuned to a
configuration where the electron wavefunctions hybridize and form
molecular states extending over both QDs.
The DQD thus provides a tunable two-level system, which has been
utilized to perform coherent manipulation of a single charge in
semiconductor nanostructures \cite{hayashi:2003,petta:2004}.

An alternative approach to molecular states at large detuning is to
study electron transport in the DQD in the framework of cotunneling
\cite{golovach:2004}.
Cotunneling involves an electron (or hole) that virtually tunnels
through an energetically forbidden charge state of the QD positioned
at an energy $\delta$ away from the Fermi energy in the leads. The
process occurs on a timescale $\tau_\mathrm{cot} \sim \hbar/\delta$
limited by time-energy uncertainty \cite{averin:1990}.
Cotunneling currents are generally small and difficult to measure,
but the effect has been utilized for QD spectroscopy
\cite{franceschi:2001,zumbuhl:2004}, for studying
cotunneling-mediated transport in single QDs
\cite{schleserPRL:2005}, or for investigating spin effects in double
QDs \cite{liu:2005}.

In this work we use a quantum point contact (QPC) as a charge sensor
\cite{field:1993} to detect single-electron tunneling in the DQD in
real-time \cite{fujisawa:2004,schleser:2004, vandersypen:2004}.
Similar setups have been used for investigating single-spin dynamics
\cite{elzermanNature:2004}, detecting single-particle interference
\cite{gustavssonNL:2008}, probing interactions between charge
carriers \cite{gustavsson:2005} or for measuring extremely small
currents \cite{fujisawa:2006, gustavssonAPL:2008}.
Here, we utilize the technique to count electrons cotunneling
through the DQD. The method provides a precise measurement of the
tunneling probability as a function of energy separation $\delta$
between the QDs, allowing a direct comparison with the rate expected
from time-energy uncertainty.
In the limit of $\delta \rightarrow 0$, the electrons form molecular
states extending over both QDs. Here, we measure tunneling rates
expected from sequential tunneling into bonding and antibonding
states of the DQD. The results experimentally establish the
equivalence between cotunneling into coupled QD states and
sequential tunneling into molecular states of the DQD.

In the last part of the paper we investigate inelastic cotunneling
processes involving excited states of the DQD. Finally, we use the
time-resolved charge-detection techniques to extract the shot noise
of the DQD current in the cotunneling regime.



\section{Experimental setup and methods}
\begin{figure}[tb]
\centering
\includegraphics[width=\linewidth]{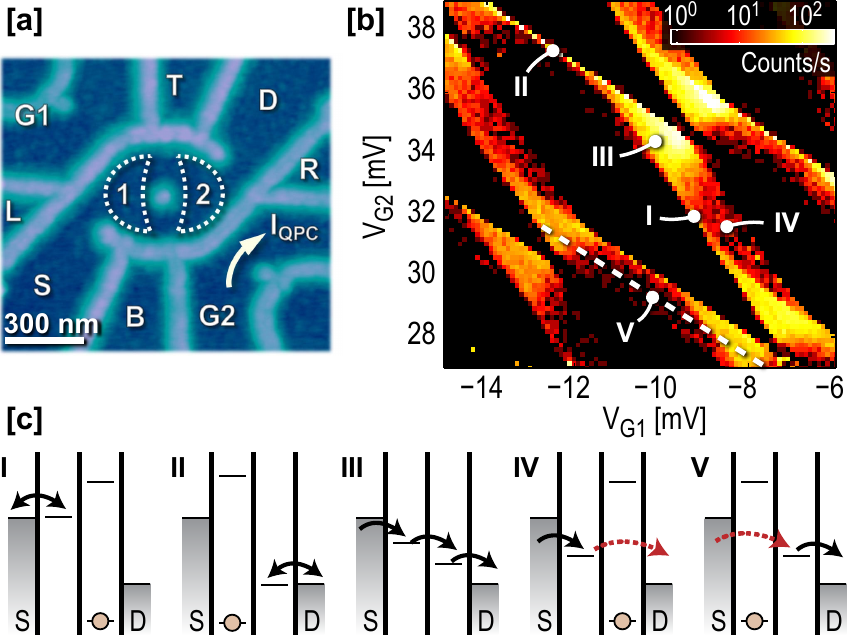}
\caption{(a) AFM-image of the sample. The structure consists of a
double quantum dot (DQD) (marked by 1 and 2) with a near-by quantum
point contact. (b) Charge stability diagram of the DQD, measured by
counting electrons entering and leaving the DQD. The data was taken
with a voltage bias of $V_\mathrm{DQD-SD} = 600\uV$ applied over the
DQD. The QPC conductance was measured with $V_\mathrm{QPC-SD} =
300\uV$. The count rates were extracted from traces of length
$T=0.5\s$. (c) Energy level diagrams for different configurations in
(b).
} \label{fig:Sample}
\end{figure}

The measurements were performed on the structure shown in
\FigRef{fig:Sample}(a). The sample is fabricated with local
oxidation \cite{fuhrer:2004} of a GaAs/Al$_{0.3}$Ga$_{0.7}$As
heterostructure, containing a two-dimensional electron gas (2DEG) 34
nm below the surface.
The structure consists of two QDs (marked by 1 and 2 in the figure)
connected by two separate tunnel barriers. Each QD contains about 30
electrons. For the results presented here only the upper tunnel
barrier was kept open; the lower was pinched-off by applying
appropriate voltages to the surrounding gates. The sample details
are described in Ref.~\cite{gustavssonPRL:2007}.

The electron population of the DQD is monitored by operating the QPC
in the lower-right corner of \FigRef{fig:Sample}(a) as a charge
detector \cite{field:1993}. By tuning the tunneling rates of the DQD
below the detector bandwidth, charge transitions can be detected in
real-time \cite{vandersypen:2004, schleser:2004, fujisawa:2004}.
In the experiment, the tunneling rates $\Gs$ and $\Gd$ to source and
drain leads are kept around 1 kHz, while the interdot coupling $t$
is set much larger ($t \sim 20\ueV \sim 5~\mathrm{GHz}$). Interdot
transitions thus occur on timescales much faster than it is possible
to register with the detector ($\tau_\mathrm{det} \sim 50 \us$)
\cite{naaman:2006}, but the coupling energy may still be determined
from charge localization measurements \cite{dicarlo:2004}.
The conductance of the QPC was measured by applying a bias voltage
of $200-400\uV$ and monitoring the current [$\Iqpc$ in
\FigRef{fig:Sample}(a)]. We ensured that the QPC bias voltage was
kept low enough to avoid charge transitions driven by current
fluctuations in the QPC \cite{gustavssonPRL:2007}. The sample is
realized without metallic gates so that the coupling between dots
and QPC detectors is not screened by metallic structures.

Figure~\ref{fig:Sample}(b) shows a charge stability diagram for the
DQD, measured by counting electrons tunneling into and out of the
DQD. The data was taken with a bias voltage of $600\uV$ applied
across the DQD, giving rise to finite-bias triangles of sequential
transport \cite{vanderwiel:2002}. The diagrams in
Fig.~\ref{fig:Sample}(c) show schematics of the DQD energy levels
for different positions in the charge stability diagram. Depending
on energy level alignment, different kinds of electron tunneling are
possible.

At the position marked by I in \FigRef{fig:Sample}(b), the
electrochemical potential $\mul$ of QD1 is aligned with the Fermi
level of the source lead. The tunneling is due to equilibrium
fluctuations between source and QD1. A measurement of the count rate
as a function of $\mul$ provides a way to determine both the
tunneling rate $\Gs$ and the electron temperature in the source lead
\cite{gustavsson:2006}. The situation is reversed at point II in
\FigRef{fig:Sample}(b). Here, electron tunneling occurs between QD2
and the drain, thus giving an independent measurement of $\Gd$ and
the electron temperature of the drain lead.
At point III within the triangle of \FigRef{fig:Sample}(b), the
levels of both QD1 and QD2 are within the bias window and the
tunneling is due to sequential transport of electrons from the
source lead into QD1, over to QD2 and finally out to the drain. The
electron flow is unidirectional and the count rate relates directly
to the current flowing through the system \cite{fujisawa:2006}.
Between the triangles, there are broad, band-shaped regions with low
but non-zero count rates where sequential transport is expected to
be suppressed due to Coulomb blockade [cases IV and V in
\FigRef{fig:Sample}(b,c)]. The finite count rate in this region is
attributed to electron tunneling involving virtual processes. These
features will be investigated in more detail in the forthcoming
sections.

\begin{figure}[b!]
\centering
\includegraphics[width=\linewidth]{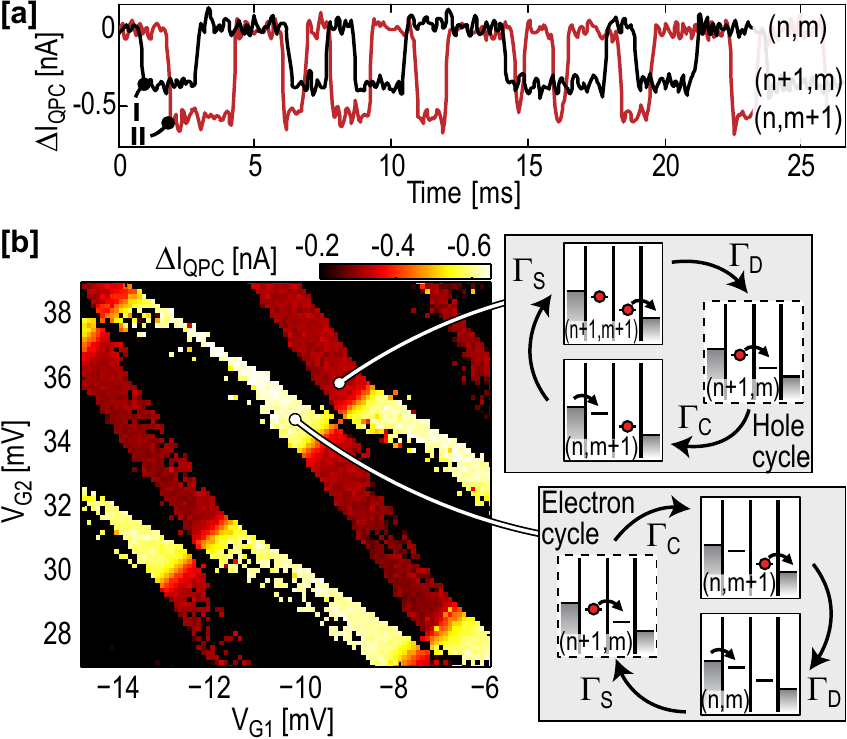}
\caption{(a) Typical time traces of the QPC current from
configurations I and II in \FigRef{fig:Sample}(b). (b) Change of QPC
current $\Delta \Iqpc$ as one electron enters the DQD, extracted
from the same set of data as shown in \FigRef{fig:Sample}. The two
levels correspond to the QPC detector registering electron tunneling
in QD1 and QD2, respectively. The energy level diagrams describe the
hole and the electron cycle of sequential transport within the
finite bias triangles.  } \label{fig:HexagonHeight}
\end{figure}

To begin with, we use the time-resolved charge detection methods to
characterize the system. Typical time traces of the QPC current for
DQD configurations marked by I and II in \FigRef{fig:Sample}(b) are
shown in \FigRef{fig:HexagonHeight}(a). The QPC current switches
between two levels, corresponding to electrons entering or leaving
QD1 (case I) or QD2 (case II). The change $\Delta \Iqpc$ as one
electron enters the DQD is larger for charge fluctuations in QD2
than in QD1. This reflects the stronger coupling between the QPC and
QD2 due to the geometry of the device. A measurement of $\Delta
\Iqpc$ thus gives information about the charge localization in the
DQD.


In \FigRef{fig:HexagonHeight}(b) we investigate the charge
localization in more detail by plotting the absolute change in QPC
current $\Delta \Iqpc$ for the same set of data as in
\FigRef{fig:Sample}(a).
The detector essentially only measures two different values of
$\Delta \Iqpc$; either $\Delta \Iqpc \sim -0.3\nA$ or $\Delta \Iqpc
\sim -0.6\nA$. Comparing the results of
\FigRef{fig:HexagonHeight}(b) with the sketches in
\FigRef{fig:Sample}(c), we see that regions with high $\Delta \Iqpc$
match with the regions where we expect the counts to be due to
electron tunneling in QD2, while the regions with low $\Delta \Iqpc$
come from electron tunneling in QD1.

The regions inside the bias triangles are described in detail in the
energy level diagrams of \FigRef{fig:HexagonHeight}(b). We assume
each QD to hold n and m electrons, respectively. In the lower
triangle, the current is carried by a sequential \emph{electron
cycle}. Starting from the (n,m)-configuration, an electron will
tunnel in from the source lead at a rate $\Gs$ making the transition
$(\mathrm{n,m})\rightarrow(\mathrm{n+1,m})$. The electron then
passes on to QD2 at a rate $\Gc \sim t/h$
$[(\mathrm{n+1,m})\rightarrow(\mathrm{n,m+1})]$ before leaving to
drain at the rate $\Gd$
$[(\mathrm{n,m+1})\rightarrow(\mathrm{n,m})]$. Since the rate $\Gc$
is much faster than the detector bandwidth (and $\Gc\gg \Gs,~\Gc\gg
\Gd$), the detector will only register transitions between the two
states $(\mathrm{n,m})$ and $(\mathrm{n,m+1})$. Therefore, we expect
the step height $\Delta \Iqpc$ within the lower triangle to be equal
to $\Delta \Iqpc$ measured for electron fluctuations in QD2, in
agreement with the results of \FigRef{fig:HexagonHeight}.

For the upper triangle, the DQD holds an additional electron and the
current is carried by a \emph{hole cycle}. Starting with both QDs
occupied $[(\mathrm{n+1,m+1})]$, an electron in QD2 may leave to the
drain $[(\mathrm{n+1,m+1})\rightarrow(\mathrm{n+1,m})]$, followed by
a fast interdot transition from QD1 to QD2
$[(\mathrm{n+1,m})\rightarrow(\mathrm{n,m+1})]$. Finally, an
electron can tunnel into QD1 from the source lead
$[(\mathrm{n,m+1})\rightarrow(\mathrm{n+1,m+1})]$. In the hole
cycle, the detector is not able to resolve the time the system stays
in the $(\mathrm{n+1,m})$ state; the measurement will only register
transitions between $(\mathrm{n+1,m+1})$ and $(\mathrm{n,m+1})$.
This corresponds to fluctuations of charge in QD1, giving the low
value of $\Delta \Iqpc$ in \FigRef{fig:HexagonHeight}(b).
Finally, we note that at the transition between regions of low and
high $\Delta \Iqpc$ the electron wavefunction delocalizes onto both
QDs. This provides a method for determining the interdot coupling
energy $t$ \cite{dicarlo:2004}. From the data in
\FigRef{fig:HexagonHeight}(b) we find tunnel couplings in the range
of $10-50\ueV$.

\section{Cotunneling}
We now focus on the regions of weak tunneling occuring in regions
outside the boundaries expected from sequential transport. In case
IV, the electrochemical potential of QD1 is within the bias window,
but the potential of QD2 is shifted below the Fermi level of the
source and not available for transport. We attribute the non-zero
count rate for this configuration to be due to electrons
\emph{cotunneling} from QD1 to the drain lead. The time-energy
uncertainty principle still allows electrons to tunnel from QD1 to
drain by means of a higher order process. In case V, the situation
is analogous but the roles of the two QDs are reversed; electrons
cotunnel from the source into QD2 and leave sequentially to the
drain lead.

\begin{figure}[tb]
\centering
\includegraphics[width=\linewidth]{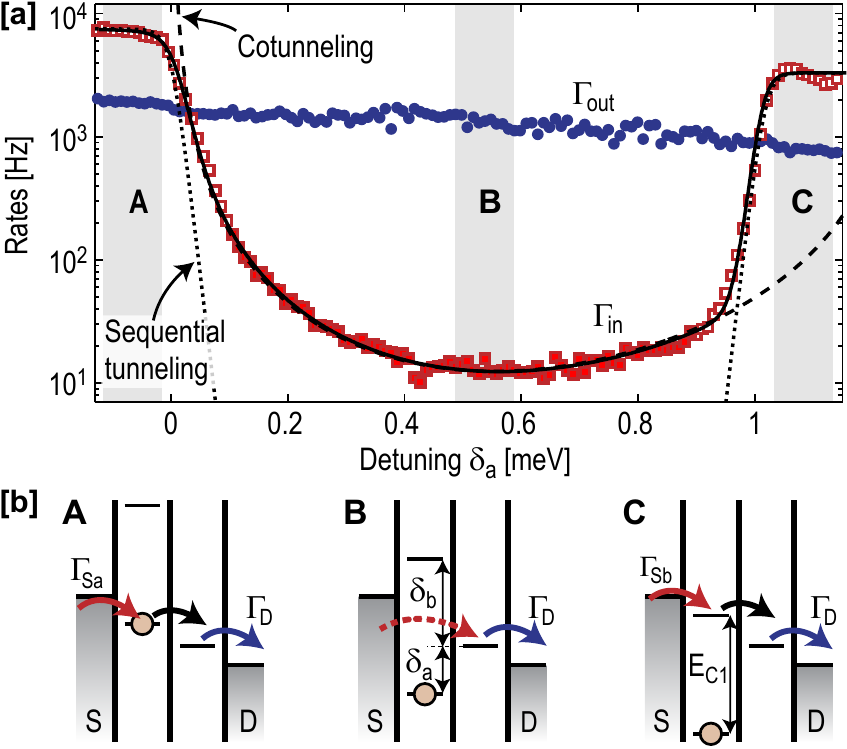}
\caption{Tunneling rates for electrons entering and leaving the DQD,
measured while keeping the potential of QD2 fixed and sweeping the
electrochemical potential of QD1. The data is measured in a
configuration similar to going along the dashed line in
\FigRef{fig:Sample}(b).
The dotted lines are tunneling rates expected from sequential
tunneling, while the dashed line is a fit to the cotunneling model
of \EqRef{eq:cotunneling}. The solid line corresponds to the model
involving molecular states [\EqRef{eq:cotMolecular}]. Parameters are
given in the text. (b) Schematic drawings of the DQD energy levels
for three different configurations in (a). At point A, electrons
tunnel sequentially through the structure. Moving to point B, the
energy levels of QD1 are shifted and the electron in QD1 is trapped
due to Coulomb blockade. Electron transport from source to QD2 is
still possible through virtual processes, but the rate for electrons
entering the DQD drops substantially due to the low probability of
the virtual processes. At point C, the next level of QD1 is brought
inside the bias window and sequential transport is again possible.}
\label{fig:CotunnelingTrace}
\end{figure}

To investigate the phenomenon more carefully, we measure the rates
for electrons tunneling into and out of the DQD in a configuration
similar to the configuration along the dashed line in
\FigRef{fig:Sample}(b). The line corresponds to keeping the
electrochemical potential of QD2 fixed within the bias window and
sweeping $\mu_1$. The data is presented in
\FigRef{fig:CotunnelingTrace}.
In the region marked by A in \FigRef{fig:CotunnelingTrace},
electrons tunnel sequentially from source into QD1, relax from QD1
down to QD2 and finally tunnel out from QD2 to the drain lead.
Proceeding from region A to region B, the electrochemical potential
$\mu_1$ is lowered so that an electron eventually gets trapped in
QD1. At point B, the electrons lack an energy $\delta_a = \mu_2 -
\mu_1$ to leave to QD2. Still, electron tunneling is possible by
means of a virtual process \cite{averin:1990}. Due to the
energy-time uncertainty principle, there is a time-window of length
$\sim\!\hbar/\delta_a$ within which tunneling from QD1 to QD2
followed by tunneling from the source into QD1 is possible without
violating energy conservation. An analogous process is possible
involving the next unoccupied state of QD1, occuring on timescales
$\sim\! \hbar/\delta_b$, where $\delta_b = E_\mathrm{C1}-\delta_a$
and $E_{C1}$ is the charging energy of QD1. The two processes
correspond to electron cotunneling from the source lead to QD2.
Continuing from point B to point C, the unoccupied state of QD2 is
shifted into the bias window and electron transport is again
sequential.

In the sequential regime (regions A and C), we fit the rate for
electrons entering the DQD to a model involving only sequential
tunneling [dotted lines in \FigRef{fig:CotunnelingTrace}(a)]
\cite{kouwenhoven:1997}. The fit allows us to determine the tunnel
couplings between source and the occupied
($\Gamma_\mathrm{Sa}$)/unoccupied ($\Gamma_\mathrm{Sb}$) states of
QD2, giving $\Gamma_\mathrm{Sa} = 7.5\kHz$, $\Gamma_\mathrm{Sb} =
3.3\kHz$ and $T = 100\mK$. Going towards region B, the rates due to
sequential tunneling are expected to drop exponentially as the
energy difference between the levels in QD1 and QD2 is increased. In
the measurement, the rate $\Gin$ initially decreases with detuning,
but the decrease is slower than exponential and flattens out as the
detuning gets larger. This is in strong disagreement with the
behavior expected for sequential tunneling. Instead, in a region
around point B we attribute the measured rate $\Gamma_\mathrm{in}$
to be due to electrons cotunneling from source to QD2.

The rate for cotunneling from source to QD2 is given as
\cite{singleCharge:1992}:
\begin{equation}\label{eq:cotunneling}
  \Gamma_\mathrm{cot} = \Gamma_\mathrm{Sa}\, \frac{t_a^2}{\delta_a^2}
  + \Gamma_\mathrm{Sb}\, \frac{t_b^2}{\delta_b^2}
  + \cos \phi \,\sqrt{\Gamma_\mathrm{Sa}\,\Gamma_\mathrm{Sb}} \, \frac{t_a\, t_b}{\delta_a\,\delta_b}.
\end{equation}
Here, $t_a$, $t_b$ are the tunnel couplings between the
occupied/unoccupied states in QD1 and the state in QD2. The first
term describes cotunneling involving the occupied state of QD1, the
second term describes the cotunneling over the unoccupied state and
the third term accounts for possible interference between the two.
The phase $\phi$ defines the phase difference between the two
processes. To determine $\phi$ one needs to be able to tune the
phases experimentally, which is not possible from the measurement
shown in \FigRef{fig:CotunnelingTrace}(a). In the following we
therefore assume the two processes to be independent ($\phi=\pi/2$).
Interference effects between cotunneling processes have been studied
in detail in Ref.~\cite{gustavssonNL:2008}.

The dashed line in \FigRef{fig:CotunnelingTrace}(a) shows the
results of \EqRef{eq:cotunneling}, with fitting parameters
$t_a=15~\mathrm{\mu eV}$ and $t_b=33~\mathrm{\mu eV}$. These values
are in good agreement with values obtained from charge localization
measurements. The values for $\Gamma_\mathrm{Sa}$ and
$\Gamma_\mathrm{Sb}$ are taken from measurements in the sequential
regimes. We emphasize that \EqRef{eq:cotunneling} is valid only if
$\delta_a, \delta_b \gg t_a, t_b$ and if sequential transport is
sufficiently suppressed. The data points used in the fitting
procedure are marked by filled squares in the figure. It should be
noted that the sequential tunneling in region C prevents
investigation of the cotunneling rate at small $\delta_b$. This can
easily be overcome by inverting the DQD bias.
The rate for electrons tunneling out of the DQD
[$\Gamma_\mathrm{out}$ in \FigRef{fig:CotunnelingTrace}(a)] shows
only slight variations over the region of interest. This is expected
since $\mu_2$ stays constant over the sweep. The slight decay of
$\Gamma_\mathrm{out}$ with increased detuning comes from tuning of
the tunnel barrier between QD2 and the drain \cite{maclean:2007}.

The cotunneling may be modified by the existence of a near-by QPC.
If the QPC were able to detect the presence electron in QD2 during
the cotunneling we would expect this to influence the cotunneling
process. For the measurements in \FigRef{fig:CotunnelingTrace}(a)
the QPC current was kept below 10 nA. This gives an average time
delay between two electrons passing the QPC of $e/I_\mathrm{QPC}
\sim \! 16~\mathrm{ps}$. Since this is larger than the typical
cotunneling time, it is unlikely that the electrons in the QPC are
capable of detecting the cotunneling process. The influence of the
QPC may become important for larger QPC currents. However, when the
QPC bias voltage is larger than the detuning ($e
V_\mathrm{QPC}>\delta$), the fluctuations in the QPC current may
start to drive inelastic charge transitions between the QDs
\cite{gustavssonPRL:2007, gustavssonNL:2008}. Such transitions will
compete with the cotunneling. For this reason it was not possible to
extract what effect the presence of the QPC may have on the
cotunneling process.

\section{Molecular states}
The overall good agreement between \EqRef{eq:cotunneling} and the
measured data demonstrates that time-resolved charge detection
techniques provide a direct way of quantitatively using the
time-energy uncertainty principle. However, a difficulty arises as
$\delta \rightarrow 0$; the cotunneling rate in
\EqRef{eq:cotunneling} diverges, as visualized for the dashed line
in \FigRef{fig:CotunnelingTrace}(a).
The problem with \EqRef{eq:cotunneling} is that it only takes
second-order tunneling processes into account. For small detuning
$\delta$ the cotunneling described in \EqRef{eq:cotunneling} must be
extended to include higher order processes \cite{pohjola:1997}.
%

\begin{figure}[b!] \centering
\includegraphics[width=\linewidth]{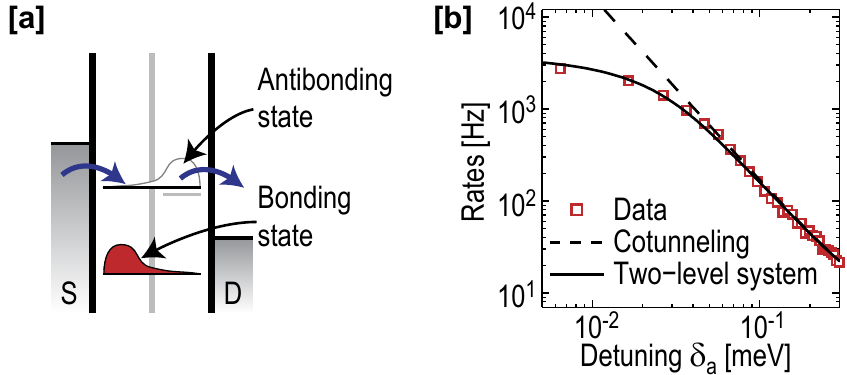}
\caption{(a) Cotunneling described using molecular states. Due to
the large detuning the empty antibonding state is mainly localized
on QD2, but a small part of the wavefunction is still present in QD1
which allows an electron to enter from the source. (b) The rate for
electrons tunneling into the DQD ($\Gin$) as a function of DQD
detuning $\delta_a$. The figure shows the same data as in
\FigRef{fig:CotunnelingTrace}, but plotted on a log-log scale to
enhance the features at small detuning. The dashed line is the
results of the cotunneling model in \EqRef{eq:cotunneling}, the
solid line shows the result of the molecular-state model
[\EqRef{eq:cotMolecular}].} \label{fig:CotunnelingTLS}
\end{figure}

A different approach is to assume the coupling between the QDs to be
fully coherent and describe the DQD in terms of the bonding and
antibonding molecular states \cite{graeber:2006,pedersen:2007}. Both
the sequential tunneling and the cotunneling can then be treated as
first-order tunneling processes into the molecular states; what we
in \FigRef{fig:CotunnelingTrace} referred to as cotunneling would be
tunneling into an antibonding state. The model is sketched in
\FigRef{fig:CotunnelingTLS}(a). The bonding state is occupied and in
Coulomb blockade. Still, an electron may tunnel from drain into the
antibonding state. Due to the large detuning, the antibonding state
is mainly located on QD2, the overlap with the electrons in the
source lead is small and the tunneling is weak. Changing the
detuning will have the effect of changing the shape of the molecular
states and shift their weights between the two QDs.

To calculate the rate for electron tunneling from source into the
molecular state of the DQD as visualized in
\FigRef{fig:CotunnelingTLS}(a), we consider the DQD as a
tunnel-coupled two-level system containing one electron, isolated
from the environment. We introduce the basis states
$\{\Psi_\mathrm{1},\Psi_\mathrm{2}\}$ describing the electron
sitting on the left or the right QD, respectively. The two states
are tunnel coupled with coupling $t$ and separated in energy by the
detuning $\delta$. The Hamiltonian of the system is
\begin{equation}\label{eq:DQ_TLSHamiltonian}
    H = \left[    \begin{array}{cc}
                   -\delta/2 & t \\
                   t & \delta/2 \\
                 \end{array}
               \right].
\end{equation}
The eigenvectors of the Hamiltonian in \EqRef{eq:DQ_TLSHamiltonian}
form the bonding $\Psi_\mathrm{B}$ and antibonding states
$\Psi_\mathrm{A}$ of the system. The eigenvalues give the energies
$E_\mathrm{B}$, $E_\mathrm{A}$ of the two states, with
\begin{equation}\label{eq:DQ_TLSEnergy}
 E_\mathrm{B} = -\frac{1}{2} \sqrt{4 t^2 + \delta^2}, ~~~~
 E_\mathrm{A} = \frac{1}{2} \sqrt{4 t^2 + \delta^2}.
\end{equation}
Note that at zero detuning there is still a finite level separation
set by the tunnel coupling. The occupation probabilities
$p_\mathrm{B}$ and $p_\mathrm{A}$ of the two states are determined
by detailed balance,
\begin{equation}\label{eq:DQ_TLSTemp}
 p_\mathrm{B} =
 1 - \frac{1}{1+e^\frac{\sqrt{4 t^2 + \delta^2}}{k_\mathrm{B}\, T}},
 ~~~
 p_\mathrm{A}  =
 \frac{1}{1+e^\frac{\sqrt{4 t^2 + \delta^2}}{k_\mathrm{B}\, T}}.
\end{equation}
To calculate the rate for electrons tunneling from source into the
antibonding molecular state of the DQD as visualized in
\FigRef{fig:CotunnelingTLS}(a), we project the thermal population
$p_\mathrm{B}$, $p_\mathrm{A}$ of the molecular states
$\Psi_\mathrm{B}$ and $\Psi_\mathrm{A}$ onto the unperturbed state
of QD1, $\Psi_1$. This gives the probability $p_1$ of finding an
electron in QD1 if making a projective measurement in the
$\Psi_1$-basis.
The measured rate $\Gin$ is equal to the probability of finding QD1
being empty $(1-p_1)$ multiplied with $\Gs$, the tunneling rate
between the source and the unperturbed state in QD1.
\begin{eqnarray}\label{eq:cotMolecular}
    \Gin &=& \Gs \, (1-p_1) = \Gs \,
    \left( 1 - (p_\mathrm{B} \Psi_\mathrm{B} + p_\mathrm{A}
    \Psi_\mathrm{A}) \cdot \Psi_1 \right) \nonumber \\
    & = & \Gs \, \frac{1}{2}
 \left(1 - \frac{\delta \tanh\left( \frac{\sqrt{4 t^2 + \delta^2}} {2k_\mathrm{B} \, T}\right)}{\sqrt{4 t^2 +
 \delta^2}} \right)
\end{eqnarray}
For large detuning, the bonding and antibonding states are well
localized in QD1 and QD2, respectively. Here, we should recover the
results for the cotunneling rate obtained for the second-order
process [\EqRef{eq:cotunneling}]. First, we assume low temperature
$k_\mathrm{B} T \ll \delta$, so that the electron only populates the
bonding ground state ($p_\mathrm{B}=1$ and $p_\mathrm{A}=0$):
\begin{equation}\label{eq:cotMolLowT}
    \Gin = \Gs \,
    \frac{1}{2} \left( 1+ \frac{\delta}{\sqrt{4t^2 +
    \delta^2}}\right).
\end{equation}
In the limit $\delta \gg t$ the relation reduces to $\Gin \approx
\Gs \, t^2/\delta^2$ and the rate approaches the result of the
second-order cotunneling processes in \EqRef{eq:cotunneling}. The
advantage of the molecular-state model is that it is valid for any
detuning, both in the sequential and in the cotunneling regime.

The solid line in \FigRef{fig:CotunnelingTrace}(a) shows the results
of \EqRef{eq:cotMolecular}. The equation has been evaluated twice,
once for the occupied [(n,m)] and once for the unoccupied state in
QD2 [(n,m+1)]; the curve in \FigRef{fig:CotunnelingTrace}(a) is the
sum of the two rates. The same parameters were used as for the
cotunneling fit of \EqRef{eq:cotunneling}. The model shows very good
agreement with data over the full range of the measurement.
To compare the results of the molecular-state and the cotunneling
model in the regime of small detuning, we plot the data in
\FigRef{fig:CotunnelingTrace}(a) on a log-log scale
[\FigRef{fig:CotunnelingTLS}(b)]. For large detuning, the tunneling
rate follows the $1/\delta^2$ predicted by both the
mole\-cular-state and the cotunneling model. For small detuning, the
deviations become apparent as the cotunneling model diverges whereas
the molecular-state model still reproduces the data well.

\section{Excited states}
\label{sec:excitedStates}
So far, we have only considered cotunneling involving the ground
states of the two QDs. The situation is more complex if we include
excited states in the model; the measured rate may come from a
combination of cotunneling processes involving different QD states.
To investigate the influence of excited states experimentally, we
start by extracting the DQD excitation spectrum using finite bias
spectroscopy \cite{vanderwiel:2002}.
If the coupling between the QDs is weak ($\tc \ll \Delta E_1,~\Delta
E_2$, with $\Delta E_{1,2}$ being the mean level spacing in each
QD), the DQD spectrum essentially consists of the combined
excitation spectrum of the individual QDs. For a more strongly
coupled DQD the QD states residing in different dots will hybridize
and delocalize over both QDs. In this section we consider a
relatively weakly coupled configuration ($t \sim 25\ueV$) and assume
the excited states to be predominantly located within the individual
QDs.


\begin{figure}[tb]
\centering
\includegraphics[width=\linewidth]{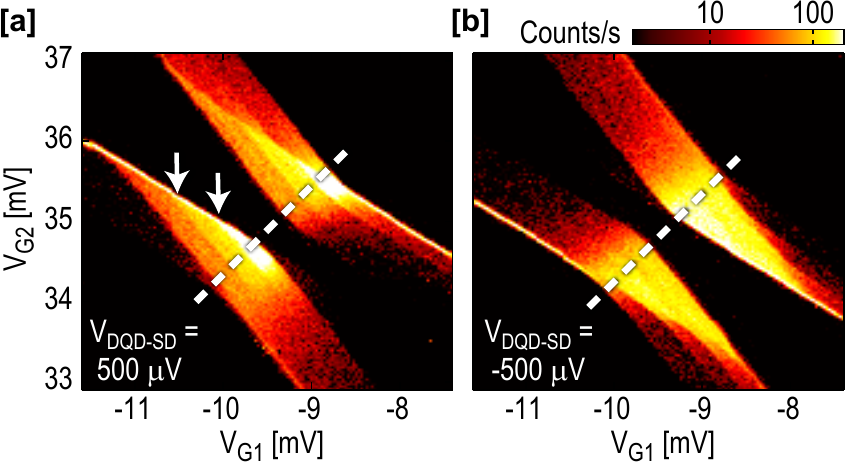}
\caption{Finite-bias spectroscopy of the DQD, taken with positive
(a) and negative (b) bias. The figures are constructed by counting
electrons entering and leaving the DQD. Excited states are visible,
especially for the positive bias data [marked with arrows in (a)].
The data was taken with $V_\mathrm{DQD-SD} = \pm 500 \uV$,
$V_\mathrm{QPC-SD} = 250 \uV$.} \label{fig:TrianglesPosNeg}
\end{figure}

\begin{figure}[b!]
\centering
\includegraphics[width=\linewidth]{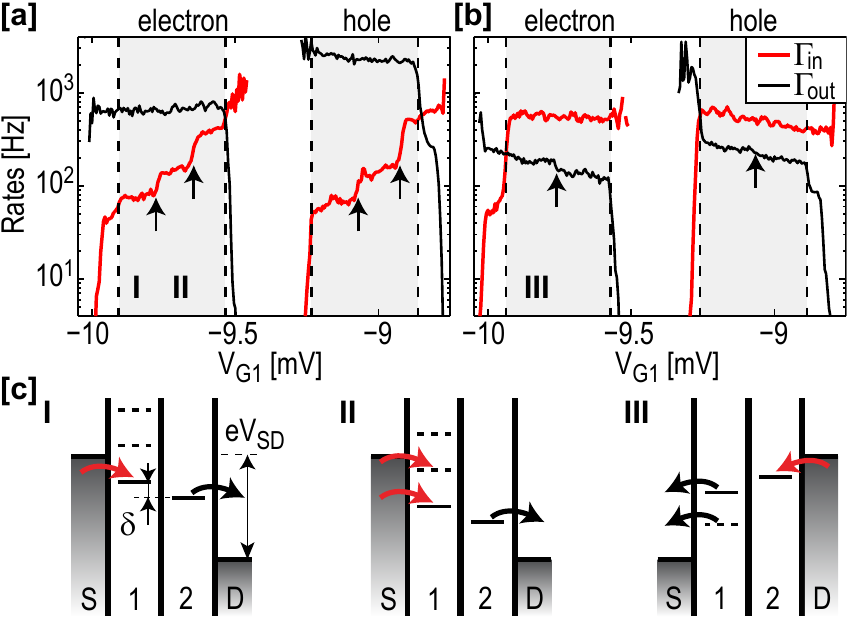}
\caption{(a,~b) Tunneling rates for electrons entering and leaving
the DQD, measured along the dashed lines in
\FigRef{fig:TrianglesPosNeg}(a,~b). In (a), we show the results for
positive bias across the DQD, in (b) the results for negative bias.
The shaded areas mark the regions where electron transport is
sequential, either in the electron or the hole transport cycle. The
arrows indicate the positions of excited states. The data was
extracted from QPC conductance traces of length $T=5\s$, taken with
$\V_\mathrm{QPC-SD} = 250\mV$. (c) Schematics of the DQD energy
configuration at three different positions in (a,~b).}
\label{fig:ExcitedStates}
\end{figure}

Figure~\ref{fig:TrianglesPosNeg} shows a magnification of two
triangles from \FigRef{fig:Sample}(b), measured with both negative
and positive bias applied across the DQD.
Excited states are visible within the triangles, especially for the
case of positive bias [marked with arrows in
\FigRef{fig:TrianglesPosNeg}(a)]. Transitions between excited states
occur along parallel lines at which the potential of QD1 is held
constant; this indicates that the excited states are located in QD1.
To investigate the states more carefully, we measure the separate
tunneling rates $\Gin$ and $\Gout$ along the dashed lines in
\FigRef{fig:TrianglesPosNeg}. The results are presented in
\FigRef{fig:ExcitedStates}, together with a few sketches depicting
the energy level configuration of the system.

We begin with the results for the positive bias case, which are
plotted in \FigRef{fig:ExcitedStates}(a). Going along the dashed
line in \FigRef{fig:TrianglesPosNeg}(a) corresponds to keeping the
detuning $\delta$ between the QDs fixed and shifting the total DQD
energy. The measurements were performed with a small detuning
($\delta \approx 100 \ueV$) to ensure that the electron transport is
unidirectional. Because of this, the outermost parts of the traces
in \FigRef{fig:ExcitedStates}(a) correspond to regions where
transport is due to cotunneling [compare the dashed line with the
position of the triangle in \FigRef{fig:TrianglesPosNeg}(a)]; the
regions where transport is sequential are shaded gray in
\FigRef{fig:ExcitedStates}(a).

Starting in the regime marked by I in
\FigRef{fig:ExcitedStates}(a,c), electrons may tunnel from source
into the ground state of QD1, relax down to QD2 and tunnel out to
the drain lead. Assuming the relaxation process to be much faster
than the other processes, the measured rates $\Gin$ and $\Gout$ are
related to the tunnel couplings of the source and drain $\Gin
\approx \Gs$ and $\Gout = \Gd$. Going to higher gate voltages lowers
the overall energy of both QDs. At the position marked by an arrow
in \FigRef{fig:ExcitedStates}(a), there is a sharp increase in the
rate for tunneling into the DQD. We attribute this to the existence
of an excited state in QD1; as shown in case II in
\FigRef{fig:ExcitedStates}(c), the electron tunneling from source
into QD1 may enter either into the ground $\mathrm{(n+1,m)}$ or the
excited state $\mathrm{(n+1^*,m)}$, giving an increase in $\Gin$.
When further lowering the DQD energy another excited state comes
into the bias window and $\Gin$ increases even more [second arrow in
\FigRef{fig:ExcitedStates}(a)]. The rate for tunneling out of the
DQD shows only minor variations within the region of interest. This
supports the assumption that the excited states quickly relax and
that the electron tunnels out of the DQD from the ground state of
QD2

Finally, continuing to the edge of the shaded region ($\Vgl \sim
-9.55\mV$), the potential of QD2 goes below the Fermi level of the
drain. Here, electrons get trapped in QD2 and the tunneling-out rate
drops drastically. At the same time, $\Gin$ increases; when the
electron in QD2 eventually tunnels out, the DQD may be refilled from
either the source or the drain lead.
The picture described above is repeated in the triangle with hole
transport ($-9.25\mV < \Vgr < -8.9\mV$). This is expected, since the
hole transport cycle involves the same QD states as in the electron
case. An interesting feature is that $\Gin$ shows essentially the
same values in both the electron and the hole cycle, while $\Gout$
increases by a factor of three. The presence of the additional
electron in QD1 apparently affects the tunnel barrier between drain
and QD2 more than an additional electron in QD2 affects the barrier
between QD1 and source.

Next, we move over to the case of negative bias
[\FigRef{fig:ExcitedStates}(b)]. Here, the roles of QD1 and QD2 are
inverted, meaning that electrons enter the DQD into QD2 and leave
from QD1. Following the data and the arguments presented for the
case of positive bias, one would expect this configuration to be
suitable for detecting excited states in QD2. However, looking at
the tunneling rates within the sequential region of
\FigRef{fig:ExcitedStates}(b), the rate for entering QD2 ($\Gin$)
stays fairly constant, while the rate for tunneling out decreases at
the point marked by the arrow. Again, we attribute the behavior to
the existence of an excited state in QD1.

The situation is described in sketch III of
\FigRef{fig:ExcitedStates}(c). The electrochemical potential of QD1
is high enough to allow the electron in the $\mathrm{(n+1,m)}$-state
to tunnel out to the source and leave the DQD in an excited state
$\mathrm{(n^*,m)}$. Since the energy difference
$E[\mathrm{(n^*,m)}]-E[\mathrm{(n+1,m)}]$ is smaller than
$E[\mathrm{(n,m)}]-E[\mathrm{(n+1,m)}]$, the transition involving
the excited state appears \emph{below} the ground state transition.
As the overall DQD potential is lowered, the transition energy
involving the excited state goes below the Fermi level of the drain,
resulting in a drop of $\Gout$ as only the ground state transition
is left available. Similar to the single QD case
\cite{gustavsson:2006}, the tunneling-in rate samples the excitation
spectrum for the {(n+1,m)}-configuration, while the tunneling-out
rate reflects the excitation spectrum of the $\mathrm{(n,m)}$-DQD.

To conclude the results of \FigRef{fig:ExcitedStates}, we find two
excited states in QD1 in the $\mathrm{(n+1,m)}$ configuration with
$\Delta E_1^{\alpha}=180\ueV$ and $\Delta E_1^{\beta}=340\ueV$, and
one excited state in QD1 in the $\mathrm{(n,m)}$ configuration, with
$\Delta E_1=220\ueV$. No clear excited state is visible in QD2. This
does not necessarily mean that such states do not exist; if they are
weakly coupled to the lead they will only have a minor influence on
the measured tunneling rates. Excited states in both QDs have been
measured in other configurations; there, we find similar spectra of
excited states for both QDs.

\section{Inelastic cotunneling} \label{sec:inelastic} Next, we
investigate the cotunneling process in the presence of excited
states. Looking carefully at the lower-right regions of the
negative-bias triangles in \FigRef{fig:TrianglesPosNeg}(b), we see
that the count rates in the cotunneling regions outside the
triangles are not constant along lines of fixed detuning
(corresponds to going in a direction parallel to the dashed line).
Instead, the cotunneling regions seem to split into three parallel
bands.

\begin{figure}[tb]
\centering
\includegraphics[width=\linewidth]{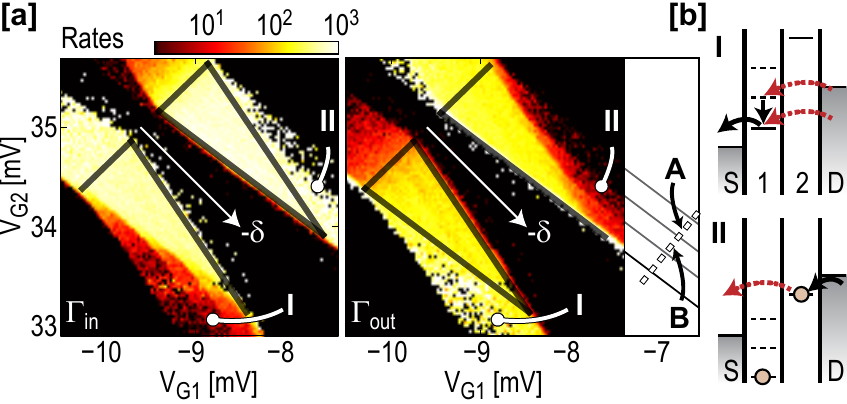}
\caption{(a) Tunneling rates for electrons entering and leaving the
DQD, extracted from the same set of data as used in
\FigRef{fig:TrianglesPosNeg}(b). The data was measured with
$V_\mathrm{DQD-SD}=-500\uV$. The solid lines mark the position of
the finite-bias triangles. The plot region in the right-hand panel
has been extended to include the regime investigated in
Fig.~\ref{fig:InelasticTrace}. (b) Energy-level diagrams for the two
positions marked in (a). In case I, the cotunneling itself is
elastic, with energy relaxation occurring after the cotunneling has
taken place. In case II, inelastic cotunneling processes are
possible.} \label{fig:InelasticMap}
\end{figure}

In \FigRef{fig:InelasticMap}(a), we plot the tunneling rates $\Gin$
and $\Gout$ for electrons entering and leaving the DQD, extracted
from the same set of data as used in
\FigRef{fig:TrianglesPosNeg}(b). The thick solid lines mark the
edges of the finite-bias triangles. Again, the cotunneling rates
outside the triangles are not uniform; parallel bands appear in
$\Gin$ for the position marked by I and in $\Gout$ for the position
marked by II in the figures.

To understand the data we draw energy level diagrams for the two
configurations [see \FigRef{fig:InelasticMap}(b)]. Focusing first on
case I, we see that the electrochemical potential of QD1 is within
the bias window, whereas QD2 is detuned and in Coulomb blockade. The
cotunneling occurs via QD2 states; electrons cotunnel from drain
into QD1, followed by sequential tunneling from QD1 to the source
lead. The picture is in agreement with what is measured in
\FigRef{fig:InelasticMap}(a); the cotunneling rate ($\Gin$) is low
and strongly depends on detuning $\delta$, while the sequential rate
$\Gout$ is high and essentially independent of detuning. The three
bands seen in $\Gin$ occur because of the excited states in QD1;
depending on the average DQD energy, electrons may cotunnel from
drain into one of the excited states, relax to the ground state and
then leave to the source lead. The state of QD2 remains unaffected
by the cotunneling process. For this configuration, we speak of
\emph{elastic} cotunneling.

The situation is different in case II. Here, cotunneling occurs in
QD1 as electrons tunnel directly from QD2 into the source lead. This
means that $\Gin$ is sequential while $\Gout$ describe the
cotunneling process. As in case I, the cotunneling rate $\Gout$
splits up into three bands; we attribute this to cotunneling where
the state of QD1 is changed during the process. QD1 ends up in one
of its excited states. The energy of the electron arriving in the
source lead is correspondingly decreased compared to the
electrochemical potential of QD2. Here, the cotunneling is
\emph{inelastic}.

\begin{figure}[b!]
\centering
\includegraphics[width=\linewidth]{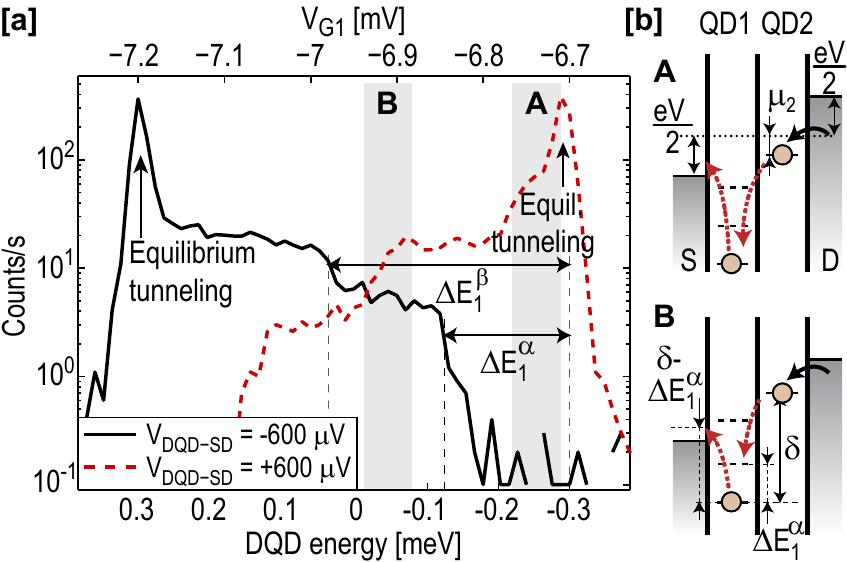}
\caption{(a) Electron count rate along the dashed line in
\FigRef{fig:InelasticMap}(a), measured for both positive and
negative DQD bias. In the trace, the detuning $\delta$ stays
constant and we sweep the average DQD energy. The DQD energy is
defined from the position where the electrochemical potential of QD2
is right in the middle between the Fermi levels of the source and
drain leads [see the dotted line in the energy level diagram in
(b)]. The steps in the count rate are due to the onset of inelastic
cotunneling processes in QD1. The data was extracted from traces of
length $T=10\s$, measured with $V_\mathrm{QPC-SD} = 200\uV$. (b)
Energy level diagrams for the two configurations marked in (a).}
\label{fig:InelasticTrace}
\end{figure}

The inelastic cotunneling is described in greater detail in
\FigRef{fig:InelasticTrace}. In \FigRef{fig:InelasticTrace}(a) we
plot the count rate for positive and negative DQD bias, measured
along the dashed line at the right edge of
\FigRef{fig:InelasticMap}(a). Figure~\ref{fig:InelasticTrace}(b)
shows energy level diagrams for negative bias at two positions along
this line. The bias voltage is applied symmetrically to the DQD,
meaning that the Fermi levels in source and drain leads are shifted
by $\pm eV/2$ relative to the Fermi energy at zero bias [dotted line
in Fig.~\ref{fig:InelasticTrace}(b)]. In the measurement of
\FigRef{fig:InelasticTrace}(a) we sweep the average DQD energy while
keeping the detuning $\delta$ constant. The average DQD energy is
defined to be zero when $\mu_2$ aligns with the zero-bias Fermi
energy in the leads [i.e. when $\mu_2 =
(\mu_\mathrm{S}+\mu_\mathrm{D})/2$].

Starting in the configuration marked by A, cotunneling is only
possible involving the QD2 ground state. Cotunneling is weak, with
count rates being well below 1 count/s. Continuing to case B, we
raise the average DQD energy. When the electrochemical potential of
QD2 is sufficiently increased compared to the Fermi level of the
source, inelastic cotunneling becomes possible leading to a sharp
increase in count rate. The process is sketched in
\FigRef{fig:InelasticTrace}(b); it involves the simultaneous
tunneling of an electron from QD2 to the first excited state of QD1
with an electron in the QD1 ground state leaving to the source. The
process is only possible if
\begin{equation}\label{eq:CondInel}
    \delta-\Delta E^\alpha_1 = \mu_1 - \mu_2 - \Delta E^\alpha_1 > \mu_\mathrm{S} - \mu_1.
\end{equation}
Here, $\Delta E^\alpha_1$ is the energy of the first excited state
in QD1. The position of the step in \FigRef{fig:InelasticTrace}(a)
directly gives the energy of the first excited state, and we find
$\Delta E^\alpha_1=180\ueV$.

Further increasing the average DQD energy makes an inelastic process
involving the second excited state in QD2 possible, giving $\Delta
E^\alpha_2=340\ueV$. Finally, as the DQD energy is raised to become
equal to half the applied bias, the electrochemical potential of QD2
aligns with Fermi level of the drain lead. Here electron tunneling
mainly occurs due to equilibrium fluctuations between drain and QD2,
giving a sharp peak in the count rate. The excited states energies
extracted from the inelastic cotunneling give the same values as
obtained from finite-bias spectroscopy within the triangles, as
described in the previous section. The good agreement between the
two measurements demonstrates the consistency of the model.

The dashed line in \FigRef{fig:InelasticTrace}(a) shows data taken
with reversed DQD bias; for this configuration the Fermi levels of
the source and drain leads are inverted, the electrons cotunnel from
source to QD2 and the peak due to equilibrium tunneling occurs at
$\mu_2= \mu_\mathrm{D} = -300\ueV$.

\section{Noise in the cotunneling regime}
Using time-resolved charge detection methods, we can extract the
noise of electron transport in the cotunneling regime. For a
weakly-coupled single QD in the regime of sequential tunneling,
transport in most configurations is well-described by independent
tunneling events for electrons entering and leaving the QD
\cite{gustavsson:2005}. The Fano factor becomes a function of the
tunneling rates \cite{davies:1992}:
\begin{equation}
F_2 = \frac{S_I}{2eI}
 = \frac{\Gamma_{\mathrm{in}}^2+\Gamma_{\mathrm{out}}^2}{(\Gamma_{\mathrm{in}}
+ \Gamma_{\mathrm{out}})^2} = \frac{1}{2} \left( 1 +a^2 \right),
\label{eq:mu2vsasym}
\end{equation}
with $a=(\Gamma_{\mathrm{in}} -
\Gamma_{\mathrm{out}})/(\Gamma_{\mathrm{in}} +
\Gamma_{\mathrm{out}})$. For symmetric barriers ($a=0$), the Fano
factor is reduced to 0.5 because of an increase in electron
correlation due to Coulomb blockade. In the case of cotunneling, the
situation is more complex. As described in the previous section,
cotunneling may involve processes leaving either QD in an excited
state. The excited state has a finite lifetime $\tau_\mathrm{rel}$;
during this time, the tunneling rates may be different compared to
the ground-state configuration \cite{schleserPRL:2005}. We therefore
expect that the existence of an electron in an excited state may
induce temporal correlations on time scales on the order of
$\tau_\mathrm{rel}$ between subsequent cotunneling events. In this
way, the noise of the cotunneling current has been proposed as a
tool to probe excited states and relaxation processes in QDs
\cite{aghassi:2006, aghassi:2008}.

\begin{figure}[t]
\centering
\includegraphics[width=\linewidth]{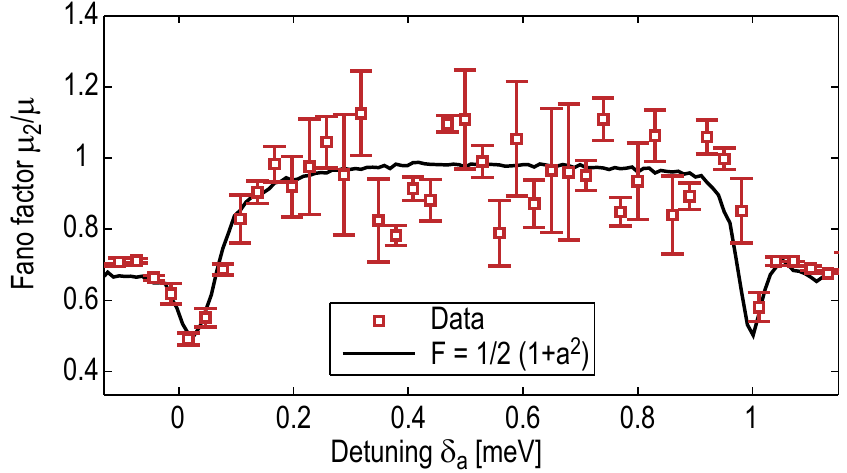}
\caption{Fano factor for electron transport in the cotunneling
regime of \FigRef{fig:CotunnelingTrace}. The data was extracted from
traces of length $T=30\s$. The solid line is the result of
\EqRef{eq:mu2vsasym}, which assumes independent tunneling events.
The minima in Fano factor occur at positions where the tunneling
rates $\Gin$ and $\Gout$ are equal (see
\FigRef{fig:CotunnelingTrace}). The error bars show standard error,
extracted by splitting the data into six subsets of length $T=5\s$
and evaluating the noise for each subset. }
\label{fig:CotunnelingNoise}
\end{figure}

In \FigRef{fig:CotunnelingNoise}, we plot the Fano factor measured
from the same region as that of \FigRef{fig:CotunnelingTrace}. The
Fano factor was extracted by measuring the distribution function for
transmitted charge through the system \cite{gustavsson:2005}. The
solid line shows the result of \EqRef{eq:mu2vsasym}, with tunneling
rates extracted from the measured traces. In the outermost regions
of the graph, the electrons tunnel sequentially through the DQD.
Here, the Fano factor is reduced due to Coulomb blockade, similar to
the single QD case. At the edges of the cotunneling regions, the
Fano factor drops further down to $F=0.5$. This is because the
injection rate $\Gin$ drops drastically as sequential transport
becomes unavailable, while $\Gout$ stays approximately constant. At
some point we get $\Gin = \Gout$, which means that the asymmetry $a$
is zero and the Fano factor of \EqRef{eq:mu2vsasym} shows a minimum.
Further into the cotunneling region, the Fano factor approaches one
as transport essentially becomes limited by a single rate; the
cotunneling rate ($\Gin$) is two orders of magnitude smaller than
the sequential rate $\Gout$.

We do not see any major deviation from the results of
\EqRef{eq:mu2vsasym}, which is only valid assuming independent
tunneling events. We have performed similar measurements in several
inelastic and elastic cotunneling regimes, without detecting any
clear deviations from \EqRef{eq:mu2vsasym}. As it turns out, there
are two effects that make it hard to detect correlations due to the
internal QD relaxations. For the first, the correlation time is
essentially set by the relaxation time $\tau_\mathrm{rel}$, which
typically occurs on a $\sim\!10~\mathrm{ns}$ timescale. This is
several orders of magnitude smaller than a typical tunneling time of
$\sim 1/\Gin \sim 100 \ms$ \cite{fujisawa:2002}. Secondly, the slow
cotunneling rate limits the amount of experimental data available
within a reasonable measurement time. This explains the large spread
between the data points in \FigRef{fig:CotunnelingNoise} in the
cotunneling regime. We conclude that the measurement bandwidth
currently limits the possibility of examining correlations in the
cotunneling regime using time-resolved detection techniques. A
higher-bandwidth detector would solve both the above mentioned
problems. It would allow a general increase of the tunneling rates
in the system, which would both decrease the difference between
$\tau_\mathrm{cot}$ and $\tau_\mathrm{rel}$ as well as provide
faster acquisition of sufficient statistics.

To conclude, we have used time-resolved charge-detection techniques
to investigate tunneling of single electrons involving virtual
processes. The measurement method provides precise determination of
all coupling energies, which allows a direct comparison with
tunneling rates expected from time-energy uncertainty. The results
give experimental confirmation of the equivalence between
cotunneling through atomic states and sequential tunneling into
molecular states. In the high-bias regime, we measure inelastic
cotunneling due to virtual processes involving excited states of the
double quantum dot. For future experiments with a high-bandwidth
detector, the method may provide a way to probe relaxation processes
and internal charge dynamics in quantum dots.

\bibliographystyle{apsrev}
\bibliography{Cotunneling}

\begin{thebibliography}{33}
\expandafter\ifx\csname natexlab\endcsname\relax\def\natexlab#1{#1}\fi
\expandafter\ifx\csname bibnamefont\endcsname\relax
  \def\bibnamefont#1{#1}\fi
\expandafter\ifx\csname bibfnamefont\endcsname\relax
  \def\bibfnamefont#1{#1}\fi
\expandafter\ifx\csname citenamefont\endcsname\relax
  \def\citenamefont#1{#1}\fi
\expandafter\ifx\csname url\endcsname\relax
  \def\url#1{\texttt{#1}}\fi
\expandafter\ifx\csname urlprefix\endcsname\relax\def\urlprefix{URL }\fi
\providecommand{\bibinfo}[2]{#2}
\providecommand{\eprint}[2][]{\url{#2}}

\bibitem[{\citenamefont{van~der Wiel et~al.}(2002)\citenamefont{van~der Wiel,
  De~Franceschi, Elzerman, Fujisawa, Tarucha, and
  Kouwenhoven}}]{vanderwiel:2002}
\bibinfo{author}{\bibfnamefont{W.~G.} \bibnamefont{van~der Wiel}},
  \bibinfo{author}{\bibfnamefont{S.}~\bibnamefont{De~Franceschi}},
  \bibinfo{author}{\bibfnamefont{J.~M.} \bibnamefont{Elzerman}},
  \bibinfo{author}{\bibfnamefont{T.}~\bibnamefont{Fujisawa}},
  \bibinfo{author}{\bibfnamefont{S.}~\bibnamefont{Tarucha}}, \bibnamefont{and}
  \bibinfo{author}{\bibfnamefont{L.~P.} \bibnamefont{Kouwenhoven}},
  \bibinfo{journal}{Rev. Mod. Phys.} \textbf{\bibinfo{volume}{75}},
  \bibinfo{pages}{1} (\bibinfo{year}{2002}).

\bibitem[{\citenamefont{Hayashi et~al.}(2003)\citenamefont{Hayashi, Fujisawa,
  Cheong, Jeong, and Hirayama}}]{hayashi:2003}
\bibinfo{author}{\bibfnamefont{T.}~\bibnamefont{Hayashi}},
  \bibinfo{author}{\bibfnamefont{T.}~\bibnamefont{Fujisawa}},
  \bibinfo{author}{\bibfnamefont{H.~D.} \bibnamefont{Cheong}},
  \bibinfo{author}{\bibfnamefont{Y.~H.} \bibnamefont{Jeong}}, \bibnamefont{and}
  \bibinfo{author}{\bibfnamefont{Y.}~\bibnamefont{Hirayama}},
  \bibinfo{journal}{Phys. Rev. Lett.} \textbf{\bibinfo{volume}{91}},
  \bibinfo{pages}{226804} (\bibinfo{year}{2003}).

\bibitem[{\citenamefont{Petta et~al.}(2004)\citenamefont{Petta, Johnson,
  Marcus, Hanson, and Gossard}}]{petta:2004}
\bibinfo{author}{\bibfnamefont{J.~R.} \bibnamefont{Petta}},
  \bibinfo{author}{\bibfnamefont{A.~C.} \bibnamefont{Johnson}},
  \bibinfo{author}{\bibfnamefont{C.~M.} \bibnamefont{Marcus}},
  \bibinfo{author}{\bibfnamefont{M.~P.} \bibnamefont{Hanson}},
  \bibnamefont{and} \bibinfo{author}{\bibfnamefont{A.~C.}
  \bibnamefont{Gossard}}, \bibinfo{journal}{Phys. Rev. Lett.}
  \textbf{\bibinfo{volume}{93}}, \bibinfo{pages}{186802}
  (\bibinfo{year}{2004}).

\bibitem[{\citenamefont{Golovach and Loss}(2004)}]{golovach:2004}
\bibinfo{author}{\bibfnamefont{V.~N.} \bibnamefont{Golovach}} \bibnamefont{and}
  \bibinfo{author}{\bibfnamefont{D.}~\bibnamefont{Loss}},
  \bibinfo{journal}{Phys. Rev. B} \textbf{\bibinfo{volume}{69}},
  \bibinfo{pages}{245327} (\bibinfo{year}{2004}).

\bibitem[{\citenamefont{Averin and Nazarov}(1990)}]{averin:1990}
\bibinfo{author}{\bibfnamefont{D.~V.} \bibnamefont{Averin}} \bibnamefont{and}
  \bibinfo{author}{\bibfnamefont{Y.~V.} \bibnamefont{Nazarov}},
  \bibinfo{journal}{Phys. Rev. Lett.} \textbf{\bibinfo{volume}{65}},
  \bibinfo{pages}{2446} (\bibinfo{year}{1990}).

\bibitem[{\citenamefont{De~Franceschi et~al.}(2001)\citenamefont{De~Franceschi,
  Sasaki, Elzerman, van~der Wiel, Tarucha, and Kouwenhoven}}]{franceschi:2001}
\bibinfo{author}{\bibfnamefont{S.}~\bibnamefont{De~Franceschi}},
  \bibinfo{author}{\bibfnamefont{S.}~\bibnamefont{Sasaki}},
  \bibinfo{author}{\bibfnamefont{J.~M.} \bibnamefont{Elzerman}},
  \bibinfo{author}{\bibfnamefont{W.~G.} \bibnamefont{van~der Wiel}},
  \bibinfo{author}{\bibfnamefont{S.}~\bibnamefont{Tarucha}}, \bibnamefont{and}
  \bibinfo{author}{\bibfnamefont{L.~P.} \bibnamefont{Kouwenhoven}},
  \bibinfo{journal}{Phys. Rev. Lett.} \textbf{\bibinfo{volume}{86}},
  \bibinfo{pages}{878} (\bibinfo{year}{2001}).

\bibitem[{\citenamefont{Zumb\"uhl et~al.}(2004)\citenamefont{Zumb\"uhl, Marcus,
  Hanson, and Gossard}}]{zumbuhl:2004}
\bibinfo{author}{\bibfnamefont{D.~M.} \bibnamefont{Zumb\"uhl}},
  \bibinfo{author}{\bibfnamefont{C.~M.} \bibnamefont{Marcus}},
  \bibinfo{author}{\bibfnamefont{M.~P.} \bibnamefont{Hanson}},
  \bibnamefont{and} \bibinfo{author}{\bibfnamefont{A.~C.}
  \bibnamefont{Gossard}}, \bibinfo{journal}{Phys. Rev. Lett.}
  \textbf{\bibinfo{volume}{93}}, \bibinfo{pages}{256801}
  (\bibinfo{year}{2004}).

\bibitem[{\citenamefont{Schleser et~al.}(2005)\citenamefont{Schleser, Ihn, Ruh,
  Ensslin, Tews, Pfannkuche, Driscoll, and Gossard}}]{schleserPRL:2005}
\bibinfo{author}{\bibfnamefont{R.}~\bibnamefont{Schleser}},
  \bibinfo{author}{\bibfnamefont{T.}~\bibnamefont{Ihn}},
  \bibinfo{author}{\bibfnamefont{E.}~\bibnamefont{Ruh}},
  \bibinfo{author}{\bibfnamefont{K.}~\bibnamefont{Ensslin}},
  \bibinfo{author}{\bibfnamefont{M.}~\bibnamefont{Tews}},
  \bibinfo{author}{\bibfnamefont{D.}~\bibnamefont{Pfannkuche}},
  \bibinfo{author}{\bibfnamefont{D.~C.} \bibnamefont{Driscoll}},
  \bibnamefont{and} \bibinfo{author}{\bibfnamefont{A.~C.}
  \bibnamefont{Gossard}}, \bibinfo{journal}{Phys. Rev. Lett.}
  \textbf{\bibinfo{volume}{94}}, \bibinfo{pages}{206805}
  (\bibinfo{year}{2005}).

\bibitem[{\citenamefont{Liu et~al.}(2005)\citenamefont{Liu, Fujisawa, Hayashi,
  and Hirayama}}]{liu:2005}
\bibinfo{author}{\bibfnamefont{H.~W.} \bibnamefont{Liu}},
  \bibinfo{author}{\bibfnamefont{T.}~\bibnamefont{Fujisawa}},
  \bibinfo{author}{\bibfnamefont{T.}~\bibnamefont{Hayashi}}, \bibnamefont{and}
  \bibinfo{author}{\bibfnamefont{Y.}~\bibnamefont{Hirayama}},
  \bibinfo{journal}{Phys. Rev. B} \textbf{\bibinfo{volume}{72}},
  \bibinfo{pages}{161305} (\bibinfo{year}{2005}).

\bibitem[{\citenamefont{Field et~al.}(1993)\citenamefont{Field, Smith, Pepper,
  Ritchie, Frost, Jones, and Hasko}}]{field:1993}
\bibinfo{author}{\bibfnamefont{M.}~\bibnamefont{Field}},
  \bibinfo{author}{\bibfnamefont{C.~G.} \bibnamefont{Smith}},
  \bibinfo{author}{\bibfnamefont{M.}~\bibnamefont{Pepper}},
  \bibinfo{author}{\bibfnamefont{D.~A.} \bibnamefont{Ritchie}},
  \bibinfo{author}{\bibfnamefont{J.~E.~F.} \bibnamefont{Frost}},
  \bibinfo{author}{\bibfnamefont{G.~A.~C.} \bibnamefont{Jones}},
  \bibnamefont{and} \bibinfo{author}{\bibfnamefont{D.~G.} \bibnamefont{Hasko}},
  \bibinfo{journal}{Phys. Rev. Lett.} \textbf{\bibinfo{volume}{70}},
  \bibinfo{pages}{1311} (\bibinfo{year}{1993}).

\bibitem[{\citenamefont{Fujisawa et~al.}(2004)\citenamefont{Fujisawa, Hayashi,
  Hirayama, Cheong, and Jeong}}]{fujisawa:2004}
\bibinfo{author}{\bibfnamefont{T.}~\bibnamefont{Fujisawa}},
  \bibinfo{author}{\bibfnamefont{T.}~\bibnamefont{Hayashi}},
  \bibinfo{author}{\bibfnamefont{Y.}~\bibnamefont{Hirayama}},
  \bibinfo{author}{\bibfnamefont{H.~D.} \bibnamefont{Cheong}},
  \bibnamefont{and} \bibinfo{author}{\bibfnamefont{Y.~H.} \bibnamefont{Jeong}},
  \bibinfo{journal}{Appl. Phys. Lett.} \textbf{\bibinfo{volume}{84}},
  \bibinfo{pages}{2343} (\bibinfo{year}{2004}).

\bibitem[{\citenamefont{Schleser et~al.}(2004)\citenamefont{Schleser, Ruh, Ihn,
  Ensslin, Driscoll, and Gossard}}]{schleser:2004}
\bibinfo{author}{\bibfnamefont{R.}~\bibnamefont{Schleser}},
  \bibinfo{author}{\bibfnamefont{E.}~\bibnamefont{Ruh}},
  \bibinfo{author}{\bibfnamefont{T.}~\bibnamefont{Ihn}},
  \bibinfo{author}{\bibfnamefont{K.}~\bibnamefont{Ensslin}},
  \bibinfo{author}{\bibfnamefont{D.~C.} \bibnamefont{Driscoll}},
  \bibnamefont{and} \bibinfo{author}{\bibfnamefont{A.~C.}
  \bibnamefont{Gossard}}, \bibinfo{journal}{Appl. Phys. Lett.}
  \textbf{\bibinfo{volume}{85}}, \bibinfo{pages}{2005} (\bibinfo{year}{2004}).

\bibitem[{\citenamefont{Vandersypen et~al.}(2004)\citenamefont{Vandersypen,
  Elzerman, Schouten, Willems~van Beveren, Hanson, and
  Kouwenhoven}}]{vandersypen:2004}
\bibinfo{author}{\bibfnamefont{L.~M.~K.} \bibnamefont{Vandersypen}},
  \bibinfo{author}{\bibfnamefont{J.~M.} \bibnamefont{Elzerman}},
  \bibinfo{author}{\bibfnamefont{R.~N.} \bibnamefont{Schouten}},
  \bibinfo{author}{\bibfnamefont{L.~H.} \bibnamefont{Willems~van Beveren}},
  \bibinfo{author}{\bibfnamefont{R.}~\bibnamefont{Hanson}}, \bibnamefont{and}
  \bibinfo{author}{\bibfnamefont{L.~P.} \bibnamefont{Kouwenhoven}},
  \bibinfo{journal}{Appl. Phys. Lett.} \textbf{\bibinfo{volume}{85}},
  \bibinfo{pages}{4394} (\bibinfo{year}{2004}).

\bibitem[{\citenamefont{Elzerman et~al.}(2004)\citenamefont{Elzerman, Hanson,
  Willems~van Beveren, Witkamp, Vandersypen, and
  Kouwenhoven}}]{elzermanNature:2004}
\bibinfo{author}{\bibfnamefont{J.~M.} \bibnamefont{Elzerman}},
  \bibinfo{author}{\bibfnamefont{R.}~\bibnamefont{Hanson}},
  \bibinfo{author}{\bibfnamefont{L.~H.} \bibnamefont{Willems~van Beveren}},
  \bibinfo{author}{\bibfnamefont{B.}~\bibnamefont{Witkamp}},
  \bibinfo{author}{\bibfnamefont{L.~M.~K.} \bibnamefont{Vandersypen}},
  \bibnamefont{and} \bibinfo{author}{\bibfnamefont{L.~P.}
  \bibnamefont{Kouwenhoven}}, \bibinfo{journal}{Nature}
  \textbf{\bibinfo{volume}{430}}, \bibinfo{pages}{431} (\bibinfo{year}{2004}).

\bibitem[{\citenamefont{Gustavsson
  et~al.}(2008{\natexlab{a}})\citenamefont{Gustavsson, Leturcq, Studer, Ihn,
  Ensslin, Driscoll, and Gossard}}]{gustavssonNL:2008}
\bibinfo{author}{\bibfnamefont{S.}~\bibnamefont{Gustavsson}},
  \bibinfo{author}{\bibfnamefont{R.}~\bibnamefont{Leturcq}},
  \bibinfo{author}{\bibfnamefont{M.}~\bibnamefont{Studer}},
  \bibinfo{author}{\bibfnamefont{T.}~\bibnamefont{Ihn}},
  \bibinfo{author}{\bibfnamefont{K.}~\bibnamefont{Ensslin}},
  \bibinfo{author}{\bibfnamefont{D.~C.} \bibnamefont{Driscoll}},
  \bibnamefont{and} \bibinfo{author}{\bibfnamefont{A.~C.}
  \bibnamefont{Gossard}}, \bibinfo{journal}{Nano Letters}
  \textbf{\bibinfo{volume}{8}}, \bibinfo{pages}{2547}
  (\bibinfo{year}{2008}{\natexlab{a}}).

\bibitem[{\citenamefont{Gustavsson
  et~al.}(2006{\natexlab{a}})\citenamefont{Gustavsson, Leturcq, Simovic,
  Schleser, Ihn, Studerus, Ensslin, Driscoll, and Gossard}}]{gustavsson:2005}
\bibinfo{author}{\bibfnamefont{S.}~\bibnamefont{Gustavsson}},
  \bibinfo{author}{\bibfnamefont{R.}~\bibnamefont{Leturcq}},
  \bibinfo{author}{\bibfnamefont{B.}~\bibnamefont{Simovic}},
  \bibinfo{author}{\bibfnamefont{R.}~\bibnamefont{Schleser}},
  \bibinfo{author}{\bibfnamefont{T.}~\bibnamefont{Ihn}},
  \bibinfo{author}{\bibfnamefont{P.}~\bibnamefont{Studerus}},
  \bibinfo{author}{\bibfnamefont{K.}~\bibnamefont{Ensslin}},
  \bibinfo{author}{\bibfnamefont{D.~C.} \bibnamefont{Driscoll}},
  \bibnamefont{and} \bibinfo{author}{\bibfnamefont{A.~C.}
  \bibnamefont{Gossard}}, \bibinfo{journal}{Phys. Rev. Lett.}
  \textbf{\bibinfo{volume}{96}}, \bibinfo{pages}{076605}
  (\bibinfo{year}{2006}{\natexlab{a}}).

\bibitem[{\citenamefont{Fujisawa et~al.}(2006)\citenamefont{Fujisawa, Hayashi,
  Tomita, and Hirayama}}]{fujisawa:2006}
\bibinfo{author}{\bibfnamefont{T.}~\bibnamefont{Fujisawa}},
  \bibinfo{author}{\bibfnamefont{T.}~\bibnamefont{Hayashi}},
  \bibinfo{author}{\bibfnamefont{R.}~\bibnamefont{Tomita}}, \bibnamefont{and}
  \bibinfo{author}{\bibfnamefont{Y.}~\bibnamefont{Hirayama}},
  \bibinfo{journal}{Science} \textbf{\bibinfo{volume}{312}},
  \bibinfo{pages}{1634} (\bibinfo{year}{2006}).

\bibitem[{\citenamefont{Gustavsson
  et~al.}(2008{\natexlab{b}})\citenamefont{Gustavsson, Shorubalko, Leturcq,
  Sch\"{o}n, and Ensslin}}]{gustavssonAPL:2008}
\bibinfo{author}{\bibfnamefont{S.}~\bibnamefont{Gustavsson}},
  \bibinfo{author}{\bibfnamefont{I.}~\bibnamefont{Shorubalko}},
  \bibinfo{author}{\bibfnamefont{R.}~\bibnamefont{Leturcq}},
  \bibinfo{author}{\bibfnamefont{S.}~\bibnamefont{Sch\"{o}n}},
  \bibnamefont{and} \bibinfo{author}{\bibfnamefont{K.}~\bibnamefont{Ensslin}},
  \bibinfo{journal}{Appl. Phys. Lett.} \textbf{\bibinfo{volume}{92}},
  \bibinfo{pages}{152101} (\bibinfo{year}{2008}{\natexlab{b}}).

\bibitem[{\citenamefont{Fuhrer et~al.}(2002)\citenamefont{Fuhrer, Dorn,
  L\"uscher, Heinzel, Ensslin, Wegscheider, and Bichler}}]{fuhrer:2004}
\bibinfo{author}{\bibfnamefont{A.}~\bibnamefont{Fuhrer}},
  \bibinfo{author}{\bibfnamefont{A.}~\bibnamefont{Dorn}},
  \bibinfo{author}{\bibfnamefont{S.}~\bibnamefont{L\"uscher}},
  \bibinfo{author}{\bibfnamefont{T.}~\bibnamefont{Heinzel}},
  \bibinfo{author}{\bibfnamefont{K.}~\bibnamefont{Ensslin}},
  \bibinfo{author}{\bibfnamefont{W.}~\bibnamefont{Wegscheider}},
  \bibnamefont{and} \bibinfo{author}{\bibfnamefont{M.}~\bibnamefont{Bichler}},
  \bibinfo{journal}{Superl. and Microstruc.} \textbf{\bibinfo{volume}{31}},
  \bibinfo{pages}{19} (\bibinfo{year}{2002}).

\bibitem[{\citenamefont{Gustavsson et~al.}(2007)\citenamefont{Gustavsson,
  Studer, Leturcq, Ihn, Ensslin, Driscoll, and Gossard}}]{gustavssonPRL:2007}
\bibinfo{author}{\bibfnamefont{S.}~\bibnamefont{Gustavsson}},
  \bibinfo{author}{\bibfnamefont{M.}~\bibnamefont{Studer}},
  \bibinfo{author}{\bibfnamefont{R.}~\bibnamefont{Leturcq}},
  \bibinfo{author}{\bibfnamefont{T.}~\bibnamefont{Ihn}},
  \bibinfo{author}{\bibfnamefont{K.}~\bibnamefont{Ensslin}},
  \bibinfo{author}{\bibfnamefont{D.~C.} \bibnamefont{Driscoll}},
  \bibnamefont{and} \bibinfo{author}{\bibfnamefont{A.~C.}
  \bibnamefont{Gossard}}, \bibinfo{journal}{Phys. Rev. Lett.}
  \textbf{\bibinfo{volume}{99}}, \bibinfo{pages}{206804}
  (\bibinfo{year}{2007}).

\bibitem[{\citenamefont{Naaman and Aumentado}(2006)}]{naaman:2006}
\bibinfo{author}{\bibfnamefont{O.}~\bibnamefont{Naaman}} \bibnamefont{and}
  \bibinfo{author}{\bibfnamefont{J.}~\bibnamefont{Aumentado}},
  \bibinfo{journal}{Phys. Rev. Lett.} \textbf{\bibinfo{volume}{96}},
  \bibinfo{pages}{100201} (\bibinfo{year}{2006}).

\bibitem[{\citenamefont{DiCarlo et~al.}(2004)\citenamefont{DiCarlo, Lynch,
  Johnson, Childress, Crockett, Marcus, Hanson, and Gossard}}]{dicarlo:2004}
\bibinfo{author}{\bibfnamefont{L.}~\bibnamefont{DiCarlo}},
  \bibinfo{author}{\bibfnamefont{H.~J.} \bibnamefont{Lynch}},
  \bibinfo{author}{\bibfnamefont{A.~C.} \bibnamefont{Johnson}},
  \bibinfo{author}{\bibfnamefont{L.~I.} \bibnamefont{Childress}},
  \bibinfo{author}{\bibfnamefont{K.}~\bibnamefont{Crockett}},
  \bibinfo{author}{\bibfnamefont{C.~M.} \bibnamefont{Marcus}},
  \bibinfo{author}{\bibfnamefont{M.~P.} \bibnamefont{Hanson}},
  \bibnamefont{and} \bibinfo{author}{\bibfnamefont{A.~C.}
  \bibnamefont{Gossard}}, \bibinfo{journal}{Phys. Rev. Lett.}
  \textbf{\bibinfo{volume}{92}}, \bibinfo{pages}{226801}
  (\bibinfo{year}{2004}).

\bibitem[{\citenamefont{Gustavsson
  et~al.}(2006{\natexlab{b}})\citenamefont{Gustavsson, Leturcq, Simovic,
  Schleser, Studerus, Ihn, Ensslin, Driscoll, and Gossard}}]{gustavsson:2006}
\bibinfo{author}{\bibfnamefont{S.}~\bibnamefont{Gustavsson}},
  \bibinfo{author}{\bibfnamefont{R.}~\bibnamefont{Leturcq}},
  \bibinfo{author}{\bibfnamefont{B.}~\bibnamefont{Simovic}},
  \bibinfo{author}{\bibfnamefont{R.}~\bibnamefont{Schleser}},
  \bibinfo{author}{\bibfnamefont{P.}~\bibnamefont{Studerus}},
  \bibinfo{author}{\bibfnamefont{T.}~\bibnamefont{Ihn}},
  \bibinfo{author}{\bibfnamefont{K.}~\bibnamefont{Ensslin}},
  \bibinfo{author}{\bibfnamefont{D.~C.} \bibnamefont{Driscoll}},
  \bibnamefont{and} \bibinfo{author}{\bibfnamefont{A.~C.}
  \bibnamefont{Gossard}}, \bibinfo{journal}{Phys. Rev. B}
  \textbf{\bibinfo{volume}{74}}, \bibinfo{pages}{195305}
  (\bibinfo{year}{2006}{\natexlab{b}}).

\bibitem[{\citenamefont{Kouwenhoven et~al.}(1997)\citenamefont{Kouwenhoven,
  Marcus, McEuen, Tarucha, Westervelt, and Wingreen}}]{kouwenhoven:1997}
\bibinfo{author}{\bibfnamefont{L.~P.} \bibnamefont{Kouwenhoven}},
  \bibinfo{author}{\bibfnamefont{C.~M.} \bibnamefont{Marcus}},
  \bibinfo{author}{\bibfnamefont{P.~M.} \bibnamefont{McEuen}},
  \bibinfo{author}{\bibfnamefont{S.}~\bibnamefont{Tarucha}},
  \bibinfo{author}{\bibfnamefont{R.~M.} \bibnamefont{Westervelt}},
  \bibnamefont{and} \bibinfo{author}{\bibfnamefont{N.~S.}
  \bibnamefont{Wingreen}}, in \emph{\bibinfo{booktitle}{Mesoscopic Electron
  Transport}}, edited by \bibinfo{editor}{\bibfnamefont{L.~L.}
  \bibnamefont{Sohn}}, \bibinfo{editor}{\bibfnamefont{L.~P.}
  \bibnamefont{Kouwenhoven}}, \bibnamefont{and}
  \bibinfo{editor}{\bibfnamefont{G.}~\bibnamefont{Sch\"on}}
  (\bibinfo{publisher}{Kluwer}, \bibinfo{address}{Dordrecht},
  \bibinfo{year}{1997}), NATO ASI Ser. E 345, pp. \bibinfo{pages}{105--214}.

\bibitem[{\citenamefont{Averin and Nazarov}(1992)}]{singleCharge:1992}
\bibinfo{author}{\bibfnamefont{D.~V.} \bibnamefont{Averin}} \bibnamefont{and}
  \bibinfo{author}{\bibfnamefont{Y.~V.} \bibnamefont{Nazarov}},
  \emph{\bibinfo{title}{Single Charge Tunneling}} (\bibinfo{publisher}{Plenum,
  New York}, \bibinfo{year}{1992}).

\bibitem[{\citenamefont{MacLean et~al.}(2007)\citenamefont{MacLean, Amasha,
  Radu, Zumb\"{u}hl, Kastner, Hanson, and Gossard}}]{maclean:2007}
\bibinfo{author}{\bibfnamefont{K.}~\bibnamefont{MacLean}},
  \bibinfo{author}{\bibfnamefont{S.}~\bibnamefont{Amasha}},
  \bibinfo{author}{\bibfnamefont{I.~P.} \bibnamefont{Radu}},
  \bibinfo{author}{\bibfnamefont{D.~M.} \bibnamefont{Zumb\"{u}hl}},
  \bibinfo{author}{\bibfnamefont{M.~A.} \bibnamefont{Kastner}},
  \bibinfo{author}{\bibfnamefont{M.~P.} \bibnamefont{Hanson}},
  \bibnamefont{and} \bibinfo{author}{\bibfnamefont{A.~C.}
  \bibnamefont{Gossard}}, \bibinfo{journal}{Phys. Rev. Lett.}
  \textbf{\bibinfo{volume}{98}}, \bibinfo{pages}{036802}
  (\bibinfo{year}{2007}).

\bibitem[{\citenamefont{Pohjola et~al.}(1997)\citenamefont{Pohjola, Konig,
  Salomaa, Schmid, Schoeller, and Schon}}]{pohjola:1997}
\bibinfo{author}{\bibfnamefont{T.}~\bibnamefont{Pohjola}},
  \bibinfo{author}{\bibfnamefont{J.}~\bibnamefont{Konig}},
  \bibinfo{author}{\bibfnamefont{M.~M.} \bibnamefont{Salomaa}},
  \bibinfo{author}{\bibfnamefont{J.}~\bibnamefont{Schmid}},
  \bibinfo{author}{\bibfnamefont{H.}~\bibnamefont{Schoeller}},
  \bibnamefont{and} \bibinfo{author}{\bibfnamefont{G.}~\bibnamefont{Schon}},
  \bibinfo{journal}{Europhysics Letters} \textbf{\bibinfo{volume}{40}},
  \bibinfo{pages}{189} (\bibinfo{year}{1997}).

\bibitem[{\citenamefont{Gr\"{a}ber et~al.}(2006)\citenamefont{Gr\"{a}ber,
  Coish, Hoffmann, Weiss, Furer, Oberholzer, Loss, and
  Sch\"{o}nenberger}}]{graeber:2006}
\bibinfo{author}{\bibfnamefont{M.~R.} \bibnamefont{Gr\"{a}ber}},
  \bibinfo{author}{\bibfnamefont{W.~A.} \bibnamefont{Coish}},
  \bibinfo{author}{\bibfnamefont{C.}~\bibnamefont{Hoffmann}},
  \bibinfo{author}{\bibfnamefont{M.}~\bibnamefont{Weiss}},
  \bibinfo{author}{\bibfnamefont{J.}~\bibnamefont{Furer}},
  \bibinfo{author}{\bibfnamefont{S.}~\bibnamefont{Oberholzer}},
  \bibinfo{author}{\bibfnamefont{D.}~\bibnamefont{Loss}}, \bibnamefont{and}
  \bibinfo{author}{\bibfnamefont{C.}~\bibnamefont{Sch\"{o}nenberger}},
  \bibinfo{journal}{Phys. Rev. B} \textbf{\bibinfo{volume}{74}},
  \bibinfo{pages}{075427} (\bibinfo{year}{2006}).

\bibitem[{\citenamefont{Pedersen et~al.}(2007)\citenamefont{Pedersen, Lassen,
  Wacker, and Hettler}}]{pedersen:2007}
\bibinfo{author}{\bibfnamefont{J.~N.} \bibnamefont{Pedersen}},
  \bibinfo{author}{\bibfnamefont{B.}~\bibnamefont{Lassen}},
  \bibinfo{author}{\bibfnamefont{A.}~\bibnamefont{Wacker}}, \bibnamefont{and}
  \bibinfo{author}{\bibfnamefont{M.~H.} \bibnamefont{Hettler}},
  \bibinfo{journal}{Phys. Rev. B} \textbf{\bibinfo{volume}{75}},
  \bibinfo{pages}{235314} (\bibinfo{year}{2007}).

\bibitem[{\citenamefont{Davies et~al.}(1992)\citenamefont{Davies, Hyldgaard,
  Hershfield, and Wilkins}}]{davies:1992}
\bibinfo{author}{\bibfnamefont{J.~H.} \bibnamefont{Davies}},
  \bibinfo{author}{\bibfnamefont{P.}~\bibnamefont{Hyldgaard}},
  \bibinfo{author}{\bibfnamefont{S.}~\bibnamefont{Hershfield}},
  \bibnamefont{and} \bibinfo{author}{\bibfnamefont{J.~W.}
  \bibnamefont{Wilkins}}, \bibinfo{journal}{Phys. Rev. B}
  \textbf{\bibinfo{volume}{46}}, \bibinfo{pages}{9620} (\bibinfo{year}{1992}).

\bibitem[{\citenamefont{Aghassi et~al.}(2006)\citenamefont{Aghassi, Thielmann,
  Hettler, and Sch\"{o}n}}]{aghassi:2006}
\bibinfo{author}{\bibfnamefont{J.}~\bibnamefont{Aghassi}},
  \bibinfo{author}{\bibfnamefont{A.}~\bibnamefont{Thielmann}},
  \bibinfo{author}{\bibfnamefont{M.~H.} \bibnamefont{Hettler}},
  \bibnamefont{and}
  \bibinfo{author}{\bibfnamefont{G.}~\bibnamefont{Sch\"{o}n}},
  \bibinfo{journal}{Phys. Rev. B} \textbf{\bibinfo{volume}{73}},
  \bibinfo{pages}{195323} (\bibinfo{year}{2006}).

\bibitem[{\citenamefont{Aghassi et~al.}(2008)\citenamefont{Aghassi, Hettler,
  and Schon}}]{aghassi:2008}
\bibinfo{author}{\bibfnamefont{J.}~\bibnamefont{Aghassi}},
  \bibinfo{author}{\bibfnamefont{M.~H.} \bibnamefont{Hettler}},
  \bibnamefont{and} \bibinfo{author}{\bibfnamefont{G.}~\bibnamefont{Schon}},
  \bibinfo{journal}{App. Phys. Lett.} \textbf{\bibinfo{volume}{92}},
  \bibinfo{pages}{202101} (\bibinfo{year}{2008}).

\bibitem[{\citenamefont{Fujisawa et~al.}(2002)\citenamefont{Fujisawa, Austing,
  Tokura, Hirayama, and Tarucha}}]{fujisawa:2002}
\bibinfo{author}{\bibfnamefont{T.}~\bibnamefont{Fujisawa}},
  \bibinfo{author}{\bibfnamefont{D.~G.} \bibnamefont{Austing}},
  \bibinfo{author}{\bibfnamefont{Y.}~\bibnamefont{Tokura}},
  \bibinfo{author}{\bibfnamefont{Y.}~\bibnamefont{Hirayama}}, \bibnamefont{and}
  \bibinfo{author}{\bibfnamefont{S.}~\bibnamefont{Tarucha}},
  \bibinfo{journal}{Nature} \textbf{\bibinfo{volume}{419}},
  \bibinfo{pages}{278} (\bibinfo{year}{2002}).

\end{thebibliography}

\end{document}